\documentstyle[psfig]{l-aa}
\newcommand{\msun}{\,M_{\odot}}

\newcommand{\gtap}{\mathrel{\hbox{\rlap{\lower.55ex \hbox {$\sim$}}
                   \kern-.3em \raise.4ex \hbox{$>$}}}}
\newcommand{\ltap}{\mathrel{\hbox{\rlap{\lower.55ex \hbox {$\sim$}}
                   \kern-.3em \raise.4ex \hbox{$<$}}}}

\newcommand{\san}{{S}$\,$}
\newcommand{\cmsq}{{\rm cm}^{-2}}
\newcommand{\ctrs}{{\rm ctr}_{11-40}}
\newcommand{\ergs}{{\rm erg}\,{\rm s}^{-1}}
\newcommand{\ergcms}{{\rm erg}\,{\rm cm}^{-2}\,{\rm s}^{-1}}
\newcommand{\nh}{N_{\rm H}}
\newcommand{\lx}{L_{\rm X}}
\newcommand{\am}{$'$}
\newcommand{\as}{$''$}

\newcommand{\x}{X$\,$}
\newcommand{\m}{M$\,$}
\newcommand{\ngc}{NGC$\,$}
\begin{document}
%%%%%%%%%%%%%%%%%%%%%%%%%%%%%%%%%%%%%%%%%%%%%%%%%%%%%%%%%%%%%%%%%%%%% TITLE
\thesaurus{08.01.2;
           10.15.2: M~67, NGC~188;
           13.25.5}
\title{X--rays from old open clusters: M~67 and NGC~188}
%\subtitle{      }
\author{ T. Belloni\inst{1} \and F.~ Verbunt\inst{2} \and R.D.~Mathieu\inst{3}
}
\offprints{T.~Belloni  }

\institute{   Astronomical Institute ``Anton Pannekoek'', 
              University of Amsterdam and Center for High-Energy Astrophysics, 
              Kruislaan 403, NL-1098 SJ Amsterdam, The Netherlands
         \and Astronomical Institute,
              P.O.Box 80000, NL-3508 TA Utrecht, The Netherlands
         \and Department of Astronomy, University of Wisconsin,
              Madison, WI 53706, U.S.A.
                }
\date{Received 15 July 1998; Accepted 17 August 1998}   
\maketitle

%%%%%%%%%%%%%%%%%%%%%%%%%%%%%%%%%%%%%%%%%%%%%%%%%%%%%%%%%%%%%%%%%% ABSTRACT

%\special{!userdict begin /bop-hook{gsave 150 70 translate
%0 rotate /Times-Roman findfont 40 scalefont setfont
%0 0 moveto 0.9 setgray (Draft 4) show grestore}def end}

\begin{abstract}
We have observed the old open clusters \m67 and \ngc188 with the
ROSAT PSPC. In \m67 we detect
a variety of X-ray sources. The X-ray emission by
a cataclysmic variable, a single hot white dwarf, two contact binaries, 
and some RS CVn systems is as expected. The X-ray emission by two binaries
located below the subgiant branch in the Hertzsprung Russell diagram of
the cluster, by a circular binary with a cool white dwarf, and by
two eccentric binaries with $P_b\gtap 700\,$d is puzzling. 
Two members of \ngc188 are detected, including the FK$\,$Com type star
D719. Another possible FK$\,$Com type star, probably not a member of \ngc188, 
is also detected.

\end{abstract}

\section{Introduction}

An observation of a stellar cluster is an observation of a
population of stars of the same age.
X-ray observations of open clusters with various ages can thus be
used to learn how the X-ray emission of stars changes as they evolve.
It is found in young clusters that stars with spectral types
later than F emit X-rays due to rapid rotation. As these stars age,
their rotation slows down, and their X-ray emission decreases
(e.g.\ Caillault 1996, Randich et al. 1996 and references therein).
\nocite{cai96} \nocite{rsp96} 
An observation of \m67 with ROSAT discovered X-rays not only from
the cataclysmic variable for which the observation was intended, 
but also from six other member stars (Belloni et al.\ 1993, hereafter BVS).
Several of these other member stars are known to be binaries, and thus
Belloni et al.\ suggested that their X-ray emission is due to rapid 
rotation caused by tidal interaction,\nocite{bvs93}
i.e.\ that these X-ray sources are similar to the RS CVn binaries,
which are well known X-ray emitters (e.g.\ Dempsey et al.\ 1993).
\nocite{dlfs93}
In the strict definition,
the rapidly rotating star in RS CVn binaries is a subgiant.
If the Hertzsprung Russell diagram of a cluster shows a Hertzsprung
gap, i.e.\ if the turnoff mass is higher than about 2.2$\msun$, it is
less likely to contain such RS CVn systems, as the (sub-)giants in these
clusters live shorter. This has been confirmed with X-ray
observations of such clusters as \ngc752 (Belloni \&\ Verbunt 1996)
and \ngc6940 (Belloni \&\ Tagliaferri 1997).\nocite{bv96}\nocite{bt97b}
We use a less stringent definition of RS CVn binaries, which includes
chromospherically active binaries consisting of 
two main-sequence stars, i.e.\ BY Draconis binaries.

In this paper we continue our study of the X-ray emission of old open
clusters. We describe the results of a second, longer, X-ray observation 
of \m67, and of an X-ray observation of another old galactic cluster,
\ngc188.
We also reanalyse our first observation of \m67, and make use of new optical
studies of the binaries in this cluster to determine the nature of the
X-ray sources.

\m67 is an ideal target for ROSAT, due to its proximity at about 850$\,$pc
and the low interstellar absorption at $E(B-V)\simeq0.03$
(Twarog \&\ Anthony-Twarog 1989), which may be converted into a hydrogen column
$\nh\simeq1.7\times10^{20}\cmsq$ (Predehl \&\ Schmitt 1995). 
(Note that these values differ from those used by BVS.)
Its age is estimated to be about 4$\,$Gyr (Dinescu et al.\ 1995).
\ngc188 is at a larger distance of about 2$\,$kpc and is reddened more
at $E(B-V)=0.09$ (von Hippel \&\ Sarajedini 1998), or 
$\nh\simeq5.0\times10^{20}\cmsq$. \nocite{hs98}
It is thought to be about 6$\,$Gyr old (Dinescu et al.\ 1995).
\nocite{ps95}\nocite{ta89}\nocite{ddgp95}

\section{Observations and data analysis}

\subsection{\m67}

\subsubsection{X-ray data}

We observed \m67 with the ROSAT PSPC between April 25 and May 9 1993, for a
total exposure of 15771 seconds; the center of the field of view
was at 8h51m24.0s, 11$^{\circ}$49$'$48.0$''$, the center of the cluster.
Our previous PSPC observation, obtained between 15 and 19 November 1991,
had a total exposure of 10515 seconds, and was centered at 
 8h50m24.0s, 11$^{\circ}$49$'$12.0$''$, about 10$'$ offset from 
the center of the cluster (Belloni et al.\ 1993).
The data were analyzed with the EXSAS package (Zimmermann et al. 1994).
\nocite{zbb+94}
We also re-analyzed the previous pointing, both separately
and merged with the new data. Since the PSPC point-spread function is not well
known at large off-axis angles, we limited our analysis to the central 20$'$
of the instrument. We followed the standard EXSAS
detection procedure (see e.g. BVS), whose final step consists of a Maximum
Likelihood detection (Cruddace et al. 1988) in three
separate bands: Total (PHA channels 11--240, 0.1--2.4 keV), Soft (ch.\ 11--40,
0.1--0.4 keV) and Hard (ch.\ 41--240, 0.4--2.4 keV).\nocite{chs88}
We ran the detection procedure on the single observations, with a maximum 
likelihood
threshold of acceptance of 13. This value is higher than the standard
value of 10, but has been chosen on the basis of a visual inspection of the
image, which also allowed us to discard obvious spurious detections. 
The detected sources in the three bands were crosscorrelated and, 
in case of a coincidence, the values of the source with the higher
likelihood value have been selected for further use. 
This procedure led to 45 sources in the new
pointing and 22 in the old one.
For each pointing separately, the X-ray sources
were then crosscorrelated with the entries of the HST Guide Star Catalog
(Lasker et al. 1990), to determine and correct for the systematic offset
(the so-called boresight correction) between the ROSAT X-ray
coordinates and the J2000 coordinate system. \nocite{lsm+90}

The two pointings are not coaligned, but are overlapping. Thus, a lenticular
region has been exposed in both observations. After having applied the
boresight corrections to the single photons in the two observations, a 
merging was made and a new detection was performed on the lenticular region
of overlap, with the same procedure described above. 26 sources
have been detected in this way.

The three lists of X-ray sources (old, new and merged) were then 
crosscorrelated to produce the final catalog, which contains 59 sources.
When a source appeared in more than one list, the detection with the
highest maximum likelihood value has been selected.
Positions and countrates of the sources can be found in Table~1.
Notice that the countrates in Table~1 are, like the positions, those of
the best detection.
Variability of the sources is discussed below.

% From tmb@astro.uva.nl Fri Dec 19 15:36:41 1997 Edited:21,22 jan FV
\nocite{san77}\nocite{mmj93}\nocite{ggla89}
\begin{table*}
\caption[o]{Summary of PSPC detections in the field of \m67. 
For each source we give the position,
the 90\% confidence radius, the countrate
with the channel band in which it is determined (T=11-240, S=11-40, H=41-240),
and where applicable the proposed optical counterparts: star
numbers according to Sanders (1977) and Montgomery et al.
(1993), the V magnitude, B--V color and the membership probability (mostly 
after Girard et al.\ 1989), and comments. Under comments we also indicate
from which list the optical counterpart was found: L1 for binaries with
known orbital period, L2 for other members of \m67, and L3 for remaining
optical objects (for details and corresponding chance coincidence
probabilities see Sect.~2.1.2).
Information from Pasquini \&\ Belloni (1998) is marked PB (identification
or presence of Ca H\&K emission or H$_\alpha$ emission).
 For sources
1-22 the sequence number is the same as in BVS; sources 5 and 20 from BVS
are not found in our new analysis, and have been discarded. Source 11 from BVS
is split in two sources, nos.\ 11 and 23; Source 15 from BVS is split in
nos.\ 15 and 24. The other sources are newly found. Ordering of sources
1-22 and 25-61 is on declination.
}
\begin{tabular}{rllrr@{\hspace{0.1cm}}rrrrrrl}
X             &
$\alpha$(2000)&
$\delta$(2000)&
$\Delta$r     &
cts/ksec     &
             &
S\#          &
MMJ\#        &
V            &
B--V         &
Mem.         &
Comment       \\
& & & & & & & & & &     \\
\hline
1  &8h49m53.8s&12$^\circ$01\am 29\as&  16\as&      1.7$\pm$ 0.5& H&
    &    &     &    &   &\\
2  &8h50m14.1s&12$^\circ$01\am 05\as&  19\as&      1.1$\pm$ 0.4& H&
    &    &     &    &   &\\
3 &8h49m23.4s&11$^\circ$54\am 47\as&  15\as&      4.8$\pm$ 0.8& H&
 262&    &     &    &  0&L3 B=16.3\\
4 &8h51m20.8s&11$^\circ$53\am 26\as&   6\as&      4.6$\pm$ 0.6& H&
1082&6493&11.25&0.41& 99&L2 PB: H$_\alpha$em.\\
6 &8h49m54.4s&11$^\circ$53\am 13\as&   9\as&      5.2$\pm$ 0.8& T&
    &    &     &    &   &PB: QSO\\
7 &8h51m07.5s&11$^\circ$53\am 01\as&   6\as&      4.4$\pm$ 0.5& H&
1077&5451&12.47&0.66& 99& L2 PB: Ca em.; also S2224 \\
8 &8h51m13.4s&11$^\circ$51\am 39\as&   5\as&      4.7$\pm$ 0.6& H&
1063&5542&13.52&1.07& 99&L1 PB: Ca em.\\
9 &8h49m35.4s&11$^\circ$50\am 27\as&  16\as&      4.0$\pm$ 0.7& H&
    &    &     &    &   &PB: QSO(?)\\
10 &8h51m23.7s&11$^\circ$49\am 49\as&   5\as&      5.0$\pm$ 0.6& H&
1040&6488&11.52&0.88&100&L1 PB: Ca em.\\
11 &8h51m23.1s&11$^\circ$48\am 26\as&   6\as&      4.0$\pm$ 0.5& H&
1019&5748&14.38&0.73& 99&L2 BVS-11a, see \x23 PB: Ca em.\\
12 &8h50m38.5s&11$^\circ$47\am 10\as&  10\as&      2.3$\pm$ 0.5& H&
    &    &     &    &   &\\
13 &8h51m18.7s&11$^\circ$47\am 00\as&   6\as&      3.4$\pm$ 0.5& H&
 999&5643&12.60&0.78&100&L1 PB: Ca em.\\
14 &8h51m04.2s&11$^\circ$46\am 19\as&  10\as&      3.3$\pm$ 0.5& H&
 759&5392&16.17&0.75&  0&L3 PB: Ca em(?)\\
15&8h50m57.1s&11$^\circ$45\am 52\as&  12\as&      3.2$\pm$ 0.5& H
   &     &     &    &   & & BVS-15, see \x24 \\
16 &8h51m27.1s&11$^\circ$46\am 57\as&   9\as&      5.2$\pm$ 0.7& S&
    &    &     &    &   & CV\\
17 &8h51m18.1s&11$^\circ$44\am 36\as&  14\as&      0.8$\pm$ 0.3& H&
 972&5615&15.17&1.07& 42&L3 PB: Ca em.\\
18 &8h49m42.7s&11$^\circ$41\am 54\as&  21\as&      1.1$\pm$ 0.4& H&
    &    &     &    &   &\\
19 &8h49m57.8s&11$^\circ$41\am 41\as&  15\as&      1.1$\pm$ 0.4& H&
 364&    &     &    & 82&L3 B=11.22 PB: no Ca em.\\
21 &8h50m11.2s&11$^\circ$35\am 35\as&  14\as&      1.8$\pm$ 0.5& H&
    &    &     &    &   &\\
22 &8h49m57.2s&11$^\circ$34\am 53\as&   4\as&     85.2$\pm$ 3.1& T&
    &    &     &    &   &\\
23 &8h51m20.6s&11$^\circ$48\am 43\as&  12\as&      6.4$\pm$ 0.8& S&
    &    &    &     &   &BVS-11b, WD\\
24&8h50m56.3s&11$^\circ$45\am 25\as&  12\as&      1.5$\pm$ 0.3& H&
    &5295&16.28&1.13&   &L3 BVS-15\\
25 &8h51m27.7s&12$^\circ$07\am 31\as&  15\as&      4.4$\pm$ 0.7& T&
    &    & 7.77&0.5 &  0&L3 HD 75638 (F0)\\
26  &8h51m25.4s&12$^\circ$02\am 57\as&   8\as&      4.7$\pm$ 0.6& H&
1113&5808&13.59&1.08& 94&L1 PB: Ca em.\\
27  &8h51m43.1s&12$^\circ$02\am 32\as&  12\as&      2.0$\pm$ 0.4& H\\
28  &8h51m23.6s&12$^\circ$01\am 33\as&  16\as&      1.6$\pm$ 0.4& H&
1112&5780&15.03&0.81& 86&L2\\
29  &8h52m22.3s&11$^\circ$59\am 58\as&  16\as&      4.1$\pm$ 0.6& H\\
30 &8h51m57.7s&11$^\circ$59\am 11\as&  19\as&      5.5$\pm$ 0.7& H&
    &6289&18.84&1.24& &L3\\
31 &8h51m37.2s&11$^\circ$59\am 03\as&   6\as&     11.7$\pm$ 1.0& T&
1327&6507&10.93&0.85&  0&L3\\
32 &8h51m27.3s&11$^\circ$55\am 34\as&  11\as&      1.2$\pm$ 0.3& H&
    &    &     &    &   &\san1092 in 95\% box\\
33 &8h52m09.9s&11$^\circ$55\am 27\as&  10\as&      4.1$\pm$ 0.6& H&
    &6401&17.36&1.63& &L3\\
34 &8h51m04.9s&11$^\circ$55\am 24\as&  10\as&      1.1$\pm$ 0.3& H&
    &5418&20.18&1.43& &L3\\
35 &8h50m37.3s&11$^\circ$54\am 11\as&  16\as&      1.1$\pm$ 0.3& H&
 628&5074&14.50&0.38& 69&L3\\
36 &8h50m57.5s&11$^\circ$52\am 48\as&  15\as&      0.5$\pm$ 0.2& H\\
37 &8h51m22.2s&11$^\circ$52\am 43\as&   9\as&      1.3$\pm$ 0.3& H&
1072&6491&11.32&0.63& 99&L1\\
38 &8h51m20.7s&11$^\circ$52\am 10\as&  18\as&      1.6$\pm$ 0.5& H&
1070&5671&13.98&0.61& 99&L1\\
39 &8h50m59.1s&11$^\circ$51\am 39\as&  13\as&      0.5$\pm$ 0.2& H&
    &5333&20.30&0.24& &L3\\
40 &8h51m37.8s&11$^\circ$50\am 53\as&   8\as&      1.8$\pm$ 0.4& H&
1282&6027&13.34&0.80& 99&L1 AH Cnc\\
41 &8h51m19.0s&11$^\circ$50\am 08\as&  14\as&      0.5$\pm$ 0.2& H&
1045&5654&12.55&0.59&100&L1 also S2217 (V=15.7)\\
42 &8h51m08.1s&11$^\circ$49\am 54\as&   5\as&      6.0$\pm$ 0.6& H&
1042&5457&15.79&0.88&  0&L3\\
43 &8h51m49.0s&11$^\circ$49\am 45\as&  17\as&      1.5$\pm$ 0.4& H&
1270&6166&12.68&0.59& 99&L2\\
44 &8h51m02.0s&11$^\circ$49\am 28\as&  13\as&      1.0$\pm$ 0.3& H&
 775&5371&12.62&0.63& 99&L2 also \san2214 (non-member)\\
45 &8h51m27.9s&11$^\circ$49\am 20\as&  12\as&      1.3$\pm$ 0.3& H&
1036&5833&12.80&0.50&100&L1 W Uma-type\\
46 &8h51m24.3s&11$^\circ$48\am 54\as&  19\as&      1.6$\pm$ 0.4& S&
1027&5781&13.27&0.60&100&L2 S1024 in 95\% box\\
47 &8h52m16.7s&11$^\circ$48\am 27\as&  10\as&      2.5$\pm$ 0.4& H&
1601&    &14.57&1.02& 39&L3\\
48 &8h51m35.6s&11$^\circ$48\am 11\as&  10\as&      1.9$\pm$ 0.4& H\\
49 &8h50m52.2s&11$^\circ$47\am 42\as&  17\as&      0.9$\pm$ 0.3& H&
 760&5263&13.38&0.51& 99&L2\\
50 &8h51m35.9s&11$^\circ$46\am 35\as&  11\as&      0.7$\pm$ 0.2& H&
1242&5993&12.70&0.72& 99&L1\\
\end{tabular}
\end{table*}
\setcounter{table}{0}
\begin{table*}
\caption[o]{continued}
\begin{tabular}{rllrr@{\hspace{0.1cm}}rrrrrrl}
X             &
$\alpha$(2000)&
$\delta$(2000)&
$\Delta$r     &
cts/ksec     &
             &
S\#          &
MMJ\#        &
V            &
B--V         &
Mem.         &
Comment       \\
& & & & & & & & & &     \\
\hline
51 &8h52m14.3s&11$^\circ$46\am 25\as&  12\as&      2.2$\pm$ 0.4& H&
    &6441&15.81&0.97& &L3\\
52 &8h51m50.4s&11$^\circ$46\am 07\as&  12\as&      1.0$\pm$ 0.3& H&
1237&6498&10.78&0.94& 99&L1\\
53 &8h51m30.5s&11$^\circ$45\am 48\as&  17\as&      0.7$\pm$ 0.2& H&
1234&5896&12.66&0.59& 99&L1 also M5885 and M5897\\
54 &8h51m04.8s&11$^\circ$41\am 58\as&  11\as&      1.2$\pm$ 0.3& H\\
55 &8h52m07.0s&11$^\circ$41\am 04\as&  15\as&      0.9$\pm$ 0.3& H\\
56 &8h52m17.4s&11$^\circ$41\am 02\as&  17\as&      1.2$\pm$ 0.3& H&
    &6468&16.90&1.68& &L3\\
57 &8h50m44.8s&11$^\circ$37\am 50\as&  19\as&      1.0$\pm$ 0.3& H&
 727&5158&12.08&0.84&  5&L3 also M5162\\
58 &8h52m06.1s&11$^\circ$37\am 15\as&  19\as&      1.9$\pm$ 0.4& H&
1414&6363&14.65&0.69& 26&L3\\
59 &8h51m52.8s&11$^\circ$36\am 01\as&  23\as&      2.7$\pm$ 0.5& H\\
60 &8h51m32.1s&11$^\circ$34\am 03\as&  18\as&      3.2$\pm$ 0.6& H&
 924&    &     &    &  0&L3 B=15.30\\
61 &8h51m17.4s&11$^\circ$31\am 54\as&  12\as&      3.6$\pm$ 0.6& H\\
\end{tabular}
\end{table*}

It is interesting to compare the sources detected in the old pointing with
the catalog in BVS. Although the data are the same, differences arise by
a better estimate of the background map, crucial for the detection procedure.
Two BVS sources, \x5 and \x20, have not been confirmed by our new analysis,
and 4 additional sources are detected. Thanks to the improved 
statistics obtained with the new observation two sources
are now resolved into two separate sources each: \x11 of BVS into \x11
and \x23, \x15 of BVS into \x15 and \x24.

Most of the sources are detected with more photons at energies above than
below 0.5$\,$keV, but the number of photons is too small to make more
accurate statements about the hardness ratios.

\begin{figure}
\centerline{\psfig{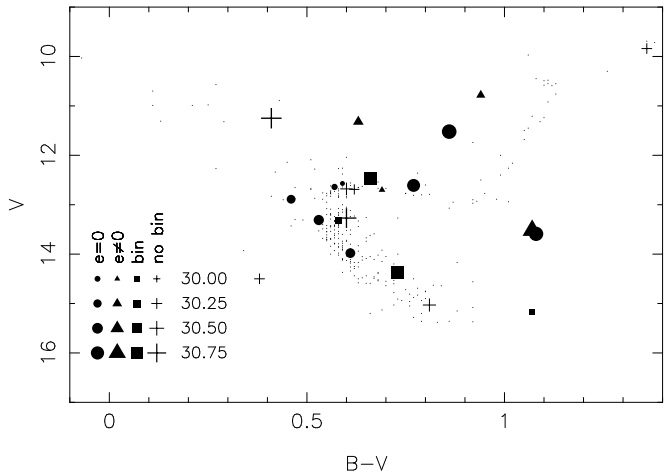} {\hfil}}
\caption[]{Hertzsprung-Russell diagram of \m67 in which the stars detected
in X-rays are indicated with special symbols. The size of a symbol is
proportional to the logarithm of the X-ray luminosity in the 0.1-2.4
keV band in erg/s. The shape of a symbol indicates its spectroscopic binary
status. Filled circles and triangles indicate circular binaries
and eccentric binaries (i.e.\ eccentricity larger than 3-$\sigma$),
respectively, squares indicate binaries with unknown period, and crosses
indicate objects for which there is no indication of binarity.
Note that for the binaries we show the total magnitude and colour.} 
\label{xray}\end{figure}

To estimate the luminosity of most sources we use a model characteristic
for chromospheric emission in close binaries, similar to the models
used by Dempsey et  al.\ (1993). Specifically we use two components
with temperatures of 0.175$\,$keV and 1.4$\,$keV, and assume that the
emission measure of the hot gas is six times higher than that of the
cool gas. 
For this model with an absorption by a column $\nh=1.7\times10^{20}\cmsq$,
1 ct/ksec in the hard band corresponds to an unabsorbed 
flux $1.8\times10^{-14}\ergcms$ in the 0.1-2.4$\,$keV band,
or a luminosity of $1.6\times10^{30}\ergs$ at the distance of \m67.

\subsubsection{Optical identifications}

The optical identification of the sources in \m67 was done in several
steps. We consider the two soft X-ray sources as having been
identified convincingly, because their proposed optical counterparts
explain their soft X-ray spectra: \x16 with the cataclysmic variable (see also
Fig.~\ref{eucnc} below), and \x23 with a white dwarf (Pasquini et al.
1994).\nocite{pba94}

We searched for further identifications in three lists: binaries in
\m67 with known periods; member stars according to Girard et al.\ (1989);
and all objects listed by Montgomery et al. (1993).
These lists are extensive and hence raise the issue of chance
coincidences. The probability of chance coincidence depends on 
location in the cluster due to the radial variation of the
stellar surface density.
Thus for each list we have considered separately an inner area with
projected distances to the cluster center $r<8'$ and an outer area
with projected distances in the range $8'<r<18'$.
The inner area corresponds to two core radii and
contains roughly 50\%\ of the cluster members
(Mathieu 1983, Mathieu \&\ Latham 1986). \nocite{mat83}
The outer boundary is chosen so as to contain all detected X-ray sources 
in the analysed area of the second ROSAT observation.

To compile the binary list 
we add to the list of binaries from Latham et al.\ (1992) the blue
stragglers from Latham \&\ Milone (1996), three contact binaries from
Gilliland et al.\ (1991), and one orbit (for \san 1113) newly determined 
by Mathieu et al.\ (1998, in preparation).
We do not include the already identified cataclysmic variable
in this part of our analysis.
\nocite{ggla89}\nocite{lmmd92}\nocite{lm96}\nocite{gbd+91}\nocite{ml86}
We obtain 11 identifications in the inner area, and 1 in the outer area,
with binaries within the 90\%\ error radius of an X-ray detection.
We estimate the probability for chance coincidences as follows.
The average error radius of an X-ray source is $0.2'$; with 36 binaries
in the inner area, the probability that one arbitrarily chosen
position is within this distance of a binary therefore is about
$36\times(0.2'/8')^2\simeq 0.02$.
The 26 X-ray sources in the inner area correspond to 26 trials, and thus
the probability for 0, 1, or 2 chance coincidences are about 55\%, 33\%,
and 10\%, respectively.
A similar reasoning for the outer area, which contains 6 of the binaries
and 22 of the X-ray sources, shows that the probability that
the one identification there is due to chance is 0.4\%.
We conclude that all 12 suggested identifications of X-ray sources
with member binaries may well be correct; but that it is also possible
that one or two identifications are spurious.

In the next step we extract from the Open Cluster DataBase of
Mermilliod (1996) \nocite{mer96} all stars in \m67 that have a
membership probability higher than 80\%\ according to Girard et al.\
(1989), and that are not included in the above-used list of
binaries. This leaves 163 stars in the inner area and 148 in the outer
area. For the remaining 15 X-ray sources in the inner area we obtain 7
new proposed counterparts; and for the remaining 21 X-ray sources in
the outer area 3 new proposed counterparts.  The probability of
getting 0, 1 or 2 chance coincidences are 20\%, 34\% and 27\%,
respectively, in the inner area; and 14\%, 29\% and 28\% in the outer
area.  
Thus at least half of the counterparts found from probable members of
\m67 in Girard et al.'s (1989) list are probably real.
Similar probabilities are indicated by the occurrence of multiple
possible identifications for some sources, such as \x7 and \x44.
In these cases we suggest the brighter optical object as the more
probable counterpart.

In the final step we compare X-ray positions with the stars in the
Tables by Montgomery et al.\ (1993). This adds a dozen other proposed 
optical counterparts for the X-ray sources, almost all of them at
faint magnitudes $V\gtap 14$;
most of these may well be chance coincidences.

All suggested optical identifications can be found in Table~1.
Many sources have no optical counterpart; clearly, considerable optical
follow-up is needed to obtain a firm knowledge of the
nature of all the X-ray sources reported in Table~1.

In the ROSAT Deep Survey, an area with 18.5$'$ radius contains about
30 sources at fluxes higher than our approximate
detection limit, $f{\rm (0.5-2.5keV)}\gtap10^{-14}\ergcms$
(see Table~4 of Hasinger et al.\ 1998). \nocite{hbg+98}
On the basis of this number we expect 6 background sources in our inner 
area, and 24 in our outer area, in the hard band (labelled H in Table~1).
Of the 26 hard X-ray sources in the inner area, we identify 11 with binaries,
and 7 with other \m67 members, a few of which may be due to chance. 
In the outer area, we identify 4 of 22 hard X-ray sources with \m67
members.
The numbers of remaining unidentified sources
are compatible with the estimated number of unrelated background sources.

Pasquini \&\ Belloni (1998) \nocite{pb98}
have obtained high- and low-resolution spectroscopy of a number of
possible optical counterparts to X-ray sources in \m67. They have
identified one, possibly two X-ray sources with with QSO's (\x6 and
\x9), and strengthened a number of our proposed identifications by
detecting Ca H\&K and/or H$_\alpha$ emission. This information is also
indicated in Table~1.

\subsection{\ngc188}

\ngc188 was observed with the ROSAT PSPC
on March 10--12 1993 for a total exposure time of 17597
seconds. The same procedure as described for \m67
was adopted to reduce the data,
with the obvious difference that only one pointing was available. The 
final catalog with 34 sources is given in Table~2. 

The two-component model for chromospheric emission discussed
above is also applied to the sources in \ngc188.
With absorption by a column $\nh=5.0\times10^{20}\cmsq$,
1 ct/ksec in the hard band corresponds to an unabsorbed 
flux of $2.1\times10^{-14}\ergcms$ in the 0.1-2.4$\,$keV band,
or a luminosity of $9.3\times10^{30}\ergs$ at the distance of \ngc188.

The catalog from the membership study by Dinescu et al. (1996) has been
used for the identifications of the X-ray sources in \ngc188. \nocite{dga+96}
The criterion for identification is the same as for \m67. 
Suggested optical identifications are listed in Table~2.
\begin{table}
\caption[o]{Summary of X-ray detections in NGC~188. The first five columns
are the same as in Table~1. For some systems we add star ID according
to Dinescu et al. (1996) and Sandage (1962), V magnitude, B--V,
membership (Dinescu et al. 1996) and comments.
}
\begin{flushleft}
\begin{tabular}{rllrrrl}
X            &
$\alpha$(2000)&
$\delta$(2000)&
$\Delta$r     &
cts/ksec      &
B         \\
\\
\hline
1 &0d47m46.6s&85$^\circ$ 35\am 13\as&11\as& 2.0$\pm$ 0.4&H\\
2 &0d36m48.1s&85$^\circ$ 35\am 09\as&19\as& 8.9$\pm$ 0.9&H\\
3 &0d38m20.1s&85$^\circ$ 34\am 27\as&23\as& 2.0$\pm$ 0.5&T\\
4 &0d36m46.4s&85$^\circ$ 32\am 23\as&14\as& 1.7$\pm$ 0.4&H\\
5 &0d44m50.0s&85$^\circ$ 32\am 01\as&12\as& 1.1$\pm$ 0.3&H\\
6 &0d39m12.4s&85$^\circ$ 31\am 26\as& 4\as&37.4$\pm$ 1.6&T\\
7 &0d42m34.5s&85$^\circ$ 31\am 21\as&14\as& 0.9$\pm$ 0.3&H\\
8 &0d46m21.9s&85$^\circ$ 30\am 53\as&12\as& 8.4$\pm$ 0.8&H\\
9 &0d37m32.6s&85$^\circ$ 30\am 14\as&16\as& 1.1$\pm$ 0.3&H\\
10 &0d42m43.5s&85$^\circ$ 29\am 07\as&15\as& 1.3$\pm$ 0.3&H\\
11 &0d36m37.3s&85$^\circ$ 28\am 59\as&15\as& 1.1$\pm$ 0.3&H\\
12 &0d37m55.9s&85$^\circ$ 28\am 46\as&10\as& 1.9$\pm$ 0.4&H\\
13 &0d44m35.2s&85$^\circ$ 27\am 07\as&12\as& 0.8$\pm$ 0.2&H\\
14 &0d39m43.5s&85$^\circ$ 26\am 38\as&13\as& 0.9$\pm$ 0.3&H\\
15 &0d33m01.0s&85$^\circ$ 24\am 52\as& 4\as&40.0$\pm$ 1.6&H\\
16 &0d51m30.7s&85$^\circ$ 24\am 46\as&13\as& 1.0$\pm$ 0.3&H\\
17 &0d45m22.0s&85$^\circ$ 23\am 40\as&15\as& 0.8$\pm$ 0.3&H\\
18 &0d37m54.6s&85$^\circ$ 22\am 54\as&11\as& 1.1$\pm$ 0.3&H\\
19 &0d50m26.3s&85$^\circ$ 22\am 06\as& 5\as& 7.6$\pm$ 0.7&H\\
20 &0d47m42.4s&85$^\circ$ 22\am 04\as& 9\as& 1.3$\pm$ 0.3&H\\
21 &0d49m58.9s&85$^\circ$ 21\am 14\as&12\as& 0.6$\pm$ 0.2&H\\
22 &0d54m16.6s&85$^\circ$ 20\am 27\as&10\as& 2.4$\pm$ 0.4&H\\
23 &0d34m18.3s&85$^\circ$ 19\am 20\as&21\as& 2.4$\pm$ 0.5&H\\
24 &0d32m29.4s&85$^\circ$ 18\am 42\as&12\as& 2.2$\pm$ 0.4&H\\
25 &0d42m59.1s&85$^\circ$ 18\am 20\as&15\as& 0.6$\pm$ 0.2&H\\
26 &0d51m22.1s&85$^\circ$ 17\am 57\as& 4\as&19.7$\pm$ 1.1&H\\
27 &0d45m27.7s&85$^\circ$ 16\am 36\as& 5\as& 5.0$\pm$ 0.6&H\\
28 &0d33m02.0s&85$^\circ$ 16\am 22\as& 6\as&11.4$\pm$ 0.9&H\\
29 &0d47m54.3s&85$^\circ$ 14\am 55\as& 8\as& 1.8$\pm$ 0.3&H\\
30 &0d42m42.7s&85$^\circ$ 14\am 15\as& 5\as& 9.7$\pm$ 0.8&H\\
31 &0d39m08.0s&85$^\circ$ 12\am 41\as&12\as& 1.2$\pm$ 0.3&H\\
32 &0d41m27.1s&85$^\circ$ 12\am 30\as&13\as& 0.9$\pm$ 0.3&H\\
33 &0d43m53.8s&85$^\circ$ 06\am 25\as&11\as& 3.9$\pm$ 0.5&H\\
34 &0d37m30.3s&85$^\circ$ 05\am 34\as& 7\as&10.1$\pm$ 0.8&H\\
\end{tabular}
\begin{tabular}{rllrrrl}
\\
\multicolumn{7}{l}{Suggested identifications; comments}\\
X            &
D\#           &
S\#           &
V             &
B--V          &
Mem.          &
Comment       \\
1 & 1824 &        &14.56&0.83&$<$60\\
2 &      &        &     &    &     &Extended?\\
5 & 1819 &        &13.98&1.20&$<$60\\
6 & 1861 &        &14.91&1.58&$<$60\\
8 &      &        &     &    &     &Extended?\\
11 & 1855 &        &12.74&0.49&$<$60\\
21 & 1335 &III-108 &15.58&0.72& 95&\\
26 & 1361 &III-89  &13.16&0.81&  1?&V8\\
29 &  719 &I-1     &11.76&1.18&100&FK Com\\
31 &      &        &     &    & \multicolumn{2}{l}{D799 in 95\% box}\\
\end{tabular}
\end{flushleft}
\end{table}

\nocite{lmmd92}\nocite{lm96}\nocite{gbd+91}

\section{Results}

\subsection{\m67}

Parameters of detected members of \m67 are given in Table~3. 
The optical identifications show that we detect stars 
spread throughout the Hertzsprung-Russell diagram of \m67, as
shown in Fig.~\ref{xray}. Accordingly, the 
types of the detected systems are varied.
Many are known photometric or spectroscopic binaries, 
including a cataclysmic variable (\x16),
two contact binaries (\x40 and \x45), three (probable) RS~CVn type systems
(\x13, \x38 and \x41), a circular binary consisting of a giant
and a white dwarf (\x10), a circular binary in a triple system
(\x53), three eccentric binaries (\x37, \x50
and \x52), two binaries located below the subgiant branch
(\x8 and \x26), and a few spectroscopic binaries for which 
there are not yet orbit solutions.
In addition we detect several sources for which there is no indication
of binarity, including a blue straggler (\x4).
The orbital periods of the binaries in \m67 detected in X-rays
range from 0.36$\,$d to 1495$\,$d.
The detection of binaries with eccentric orbits, and of some binaries with
orbital periods $P_b\gtap 40\,$d is rather surprising as no tidal
interaction is thought to take place in these (Verbunt \&\ Phinney 1995).

\begin{table}
\caption[o]{Summary of detected members (probability $>$40\%)
in \m67.
For each X-ray source we give the Sanders number, the
X-ray luminosity in $0.1-2.4\,$keV band, orbital period,
orbital eccentricity, and the binary type. All X-ray luminosities
assume an X-ray spectrum typical for chromospheric emission.
This does not apply to the white dwarf and the cataclysmic variable.
SB indicates a spectroscopic binary from the survey of Latham and
Mathieu without a determined orbit solution. 
}
\begin{flushleft}
\begin{tabular}{rrrrrl}
X            &
S\#          &
$\lx$(erg/s)   &
$P_{\rm b}$(d) &
$e$            &
comment      \\
\hline
%\multicolumn{6}{c}{M~67}\\
%\hline
4  &1082&$7.2\times 10^{30}$&     &     &blue straggler\\
7  &1077&$6.8\times 10^{30}$&SB&     & RS CVn?\\
8  &1063&$7.3\times 10^{30}$&  18.39&0.22&sub-subgiant\\
10 &1040&$7.8\times 10^{30}$&  42.83&0.00&giant$+$white dwarf\\
11 &1019&$6.2\times 10^{30}$&SB&     &RS CVn?\\
13 & 999&$5.3\times 10^{30}$&  10.06&0.00&RSCVn\\
16 & $-$ &                   &  0.09  & 0.00 &catacl.\ var.\\
17 & 972&$1.2\times 10^{30}$&SB&      &RS CVn? \\
19 & 364&$1.7\times 10^{30}$ \\
23 & $-$ &                   &       &    &white dwarf \\ 
26  &1113&$7.3\times 10^{30}$&2.82&0.03 &sub-subgiant \\
28  &1112&$2.5\times 10^{30}$&    &     &RS CVn??\\
35 & 628&$1.7\times 10^{30}$&     &     & catacl.\ var.??\\
37 &1072&$2.0\times 10^{30}$&1495.&0.32&\\
38 &1070&$2.5\times 10^{30}$&   2.66&0.00&RS CVn\\
40 &1282&$2.8\times 10^{30}$&   0.36&0.00&W Uma\\
41 &1045&$0.8\times 10^{30}$&   7.65&0.00&RS CVn\\
43 &1270&$2.3\times 10^{30}$&       &    & \\
44 & 775&$1.6\times 10^{30}$&       &    & \\
45 &1036&$2.0\times 10^{30}$&   0.44&0.00&W Uma\\
46 &1027&$6.5\times 10^{30}$&   \\
49 & 760&$1.4\times 10^{30}$&SB&      &RS CVn?\\
50 &1242&$1.1\times 10^{30}$&  31.78&0.66&\\
52 &1237&$1.6\times 10^{30}$& 697.8&0.11&\\
53 &1234&$1.1\times 10^{30}$&   4.36&0.06& triple system\\
\end{tabular}
\end{flushleft}
\end{table}

With the exception of the two contact binaries and of the cataclysmic
variable, all binaries referred to below are spectroscopic binaries.
Many of them are discussed in some detail in Mathieu et al.\ (1990).
We now proceed to discuss the sources according to category.

\subsubsection{The cataclysmic variable EU Cnc}

\begin{figure}
\centerline{\psfig{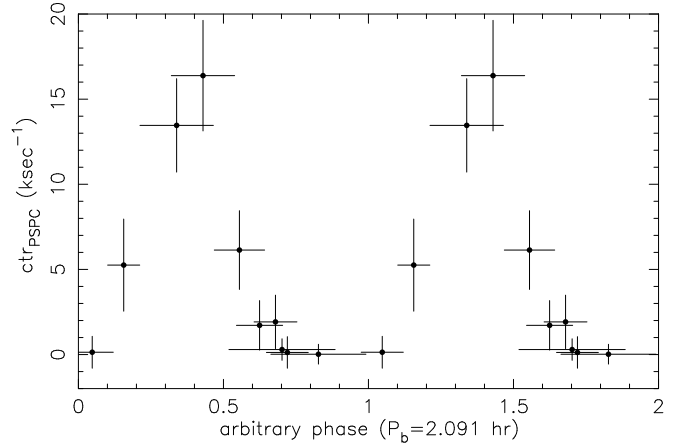} {\hfil}}
\caption[]{The folded X-ray lightcurve of EU Cnc. The 100\%\ variation
is characteristic for AM-Her type systems.} 
\label{eucnc}\end{figure}

Source \x16 is the cataclysmic variable EU Cancri. This
variable was discovered by Gilliland et al.\ (1991) \nocite{gbd+91}
at a photometric period of 2.09$\,$hr.
It is of the AM-Her type, i.e.\ a binary in which a white dwarf
with a strong magnetic field accretes from a low-mass main-sequence
dwarf, and in which the rotation of the white dwarf is locked to
the orbital revolution.
EU Cnc is detected with ROSAT only at soft energies (channels 11-40,
corresponding roughly to 0.1-0.4$\,$keV), at a countrate 
$\ctrs = 0.0052\,$cts/s.
Folding the X-ray data at the binary period, we obtain the lightcurve
shown in Fig.~\ref{eucnc}.
The predominance of soft photons is typical for AM-Her type systems,
as is the almost 100\%\ modulation of the X-ray lightcurve (due to
occultation of the X-ray emission region by the rotating white dwarf).
The ratio of $\ctrs$ to the optical flux also is in the range
observed for AM-Her type systems (see for example Table~1 of Verbunt
et al.\ 1997). \nocite{vbrp97}
Our observations thus confirm the suggestion by Gilliand et al.\ (1991)
that this object is an AM-Her type system.

Conversion of the observed countrate to flux suffers from our ignorance
of the spectrum; for an assumed blackbody spectrum with 
$kT_{\rm bb}=35-60\,$eV
we find a bolometric X-ray luminosity for EU Cnc of 
$\lx=1.2-0.7\times10^{31}\ergs$ at the distance and column of \m67.

\subsubsection{The hot white dwarf}

\begin{figure}
\centerline{\psfig{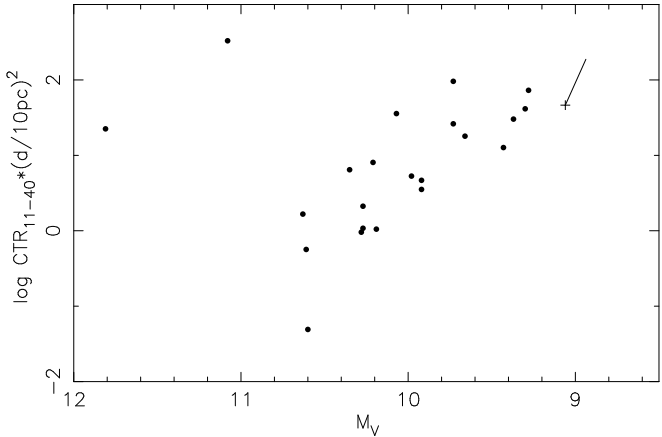} {\hfil}}
\caption[]{X-ray countrates of white dwarfs detected in the
ROSAT All Sky Survey (from Fleming et al.\ 1996), 
normalized to a distance of 10 pc, as a function
of absolute magnitude. The observed countrate for the white dwarf in 
\m67 is indicated by $+$; the correction for interstellar absorption 
depends strongly on the spectrum: the line indicates the correction
for an assumed 35 eV black body X-ray spectrum.
}
\label{wd}\end{figure}

\x23 is a very soft source, and has been identified with a white dwarf
by Pasquini et al.\ (1994). \nocite{pba94}
The effective temperature of this star is $T_{\rm eff}=68,000\pm3,000$
(Fleming et al. 1997), \nocite{flbb97}
which is hot enough to explain the X-ray emission.
To compare its countrate with that of nearby field white dwarfs,
we take the countrates found by Fleming et al.\ (1996)
in the ROSAT All Sky Survey, for white dwarfs whose absolute
magnitudes are given in the spectroscopic studies of
Bergeron et al.\ (1992) and Bragaglia et al.\ (1995).
\nocite{fsp+96}\nocite{bsl92}\nocite{brb95}
The values thus found are shown in Fig.~\ref{wd}; it is seen
that the countrate we detect for the white dwarf in \m67 is in the
range seen for field white dwarfs at comparable magnitudes.

\subsubsection{Contact binaries}

The two contact binaries detected in \m67 are AH Cnc (\x40, \san1282) and 
\san1036 (\x45, no.\ III-2 of Gilliland et al.\ 1991). \nocite{gbd+91}
The countrates are at the limit of detection. 
Contact binaries are thought to be X-ray emitters due to chromospheric 
activity, induced by the rapid rotation of the stars comprising the binary.
The X-ray luminosities of these two contact binaries may be compared with
those of contact binaries in the field, studied by McGale et al.\ (1996).
The countrates of AH~Cnc and \san1036 are between those of e.g.\ V389~Oph and
AK Her, converted to the distance of \m67.
\nocite{mph96}
\san1036 has an orbital period slightly longer than that of AK Her,
which has the longest orbital period of the systems discussed by McGale et al.
A third contact binary in \m67, III-79, was not detected; it is much
fainter in the optical than the other two, at $M_V\simeq 16$.

\subsubsection{RS CVn systems}

\begin{figure}
\centerline{\psfig{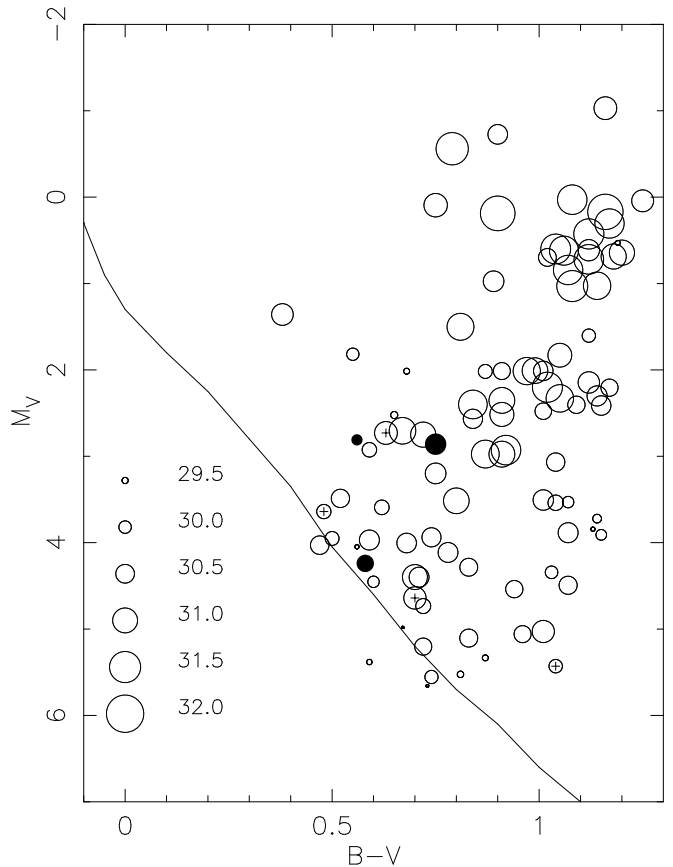} {\hfil}}
\caption[]{X-ray luminosities of RS CVn systems detected in the
ROSAT All Sky Survey (open circles; data from Dempsey et al.\ 1993) 
\nocite{dlfs93} and detected in \m67 (filled circles),
as a function of their location in the Hertzprung Russell diagram.
Colours and magnitudes are for the total light of the binary.
The size of the symbols is proportional to the logarithm of the
X-ray luminosity (in erg/s).
The solid line indicates the main sequence.
Four binaries in \m67 with unknown binary parameters presumably also
are RS CVn systems, and are also shown (circles with inscribed $+$).
Corrections for interstellar absorption have been made for the
\m67 systems only; the corrections for the field systems are expected
to be small.
The X-ray luminosities for the RS CVn systems in \m67 are similar to those 
of the field systems.
}
\label{rscvn}\end{figure}

We detect three circular binaries at orbital periods $P_b\ltap 10\,$d.
The circularity of the orbits indicates strong tidal interaction,
and we expect that these systems are chromospherically active
binaries; optical spectroscopy is required to confirm this.
\san999 and \san1045 are double-lined spectroscopic binaries discussed
by Mathieu et al. (1990).
\san999 $=$ \x13 is a binary of a slightly evolved star, hence mass
equal to the turnoff mass $1.25\msun$ of \m67, and a $1.09\msun$ 
main sequence companion.
\san1045 $=$ \x41 is a binary of two main sequence stars of 
equal masses $1.18\msun$.
\san1070 $=$ \x38 lies on the main-sequence in the Hertzsprung Russell 
diagram of \m67; its orbital period $P_b=2.66\,$d and circular orbit
suggest that it is an RS CVn binary.
The larger star in 
RS CVn binaries is forced by the tidal interaction to
corotate with the orbit, and this rapid rotation enhances the chromospheric
activity, which explains the X-ray emission.
The X-ray luminosities of the RS CVn systems in \m67 is compared with
those of RS CVn systems in the field in Fig.~\ref{rscvn}.
For the conversion of observed countrate to X-ray luminosity we use
the two-temperature model described in Sect.\ 2.1.1.
It is seen that the X-ray luminosities of the \m67 systems are 
comparable to those of RS CVn systems in similar locations of the
Hertzsprung Russell diagram.

\x53 is identified with \san1234.
Mathieu et al.\ (1990) observe that \san1234 is a triple
system in which the primary and tertiary have roughly equal light.
The period of 4.36$\,$d is then the orbital period of the inner binary.
The  visible stars would have masses close to $1.18\msun$; no strong
constraints exist for the mass of the companion in the inner binary.
The X-ray emission of \san1234 is probably due to chromospheric activity in the
inner binary: the inner orbital period is short enough for tidal interaction
to be efficient.

Four binaries with as yet unknown orbital solutions have also been detected
in X-rays (\x49 $=$ \san760, \x17 $=$ \san972, \x11 $=$ \san1019, 
\x7 $=$ \san1077). All but \san760 appear to have orbital periods of 
$\le 10$ days (for \san1077 see Mathieu et al.\ 1990), 
suggesting that they too are RS CVn systems.

\subsubsection{\san1040}

\x10 is identified with \san1040.
\san1040 was discovered to be a circular binary by Mathieu et al.\ (1990).
\nocite{mlg90}
Verbunt \&\ Phinney (1995) \nocite{vp95} showed that the giant in this binary
is too small to be responsible for the orbital circularization, and concluded
that the companion to the giant must be a white dwarf, whose progenitor
filled its Roche lobe as a giant and circularized the orbit.
The companion to the giant has been detected in the ultraviolet, and its 
ultraviolet spectrum shows that it is indeed a white dwarf, with a
temperature $T_{\rm eff}=16,160\,$K (Landsman et al.\ 1997). \nocite{lab+97}

The white dwarf is too cool to be responsible for the X-ray luminosity,
which therefore is presumably due to chromospheric activity of the
giant (Belloni et al.\ 1993). Such activity is evident from the
Mg$\,$II$\,\lambda\,$2800 doublet which is in emission 
(Landsman et al.\ 1997), and also from Ca H\&K emission
(Pasquini \&\ Belloni 1998).
We suggest that this chromospheric activity
may be a remaining effect of the
phase of mass transfer in the earlier evolution of this binary.

\subsubsection{Stars below the subgiant branch}

\x8 and \x26 are identified with \san1063 and \san1113 respectively.
This means that both stars located below the subgiant branch
in the Hertzsprung Russell diagram of \m67 have now been detected
in X-rays.
\san1063 is an eccentric binary with an orbital period of 18.3$\,$d; 
\san1113 is a circular binary with a period of 2.82$\,$d
(Latham et al.\ 1992; Mathieu et al.\ 1998, in preparation).
The nature of these stars, why they are located below the subgiant
branch, and why they emit X-rays, is a mystery. This is all the
more remarkable, as these two stars belong to the brightest X-ray
sources in \m67. 
Model computations show that the larger star in \san1063 could be 
synchronized with the orbital motion at periastron
(Van den Berg et al.\ 1998, in preparation). For $e=0.22$ this
corresponds to a rotation period which is 60\%\ of the orbital period.

The X-ray emission of \san1063 (\x8) is variable: the countrate in 
November 1991
was $8.1\pm0.9$/ksec, that in April 1993 $4.7\pm0.6$/ksec.

\subsubsection{Other eccentric binaries}

\x50 has been identified with \san1242, an eccentric binary with an
orbital period of almost 32$\,$d.
\x37 and \x52 have been identified with \san1072 and \san1237, respectively,
two long-period eccentric binaries, whose positions in the Hertzsprung
Russell diagram could suggest that they contain evolved stars
(Figs.~\ref{xray}). The colour and magnitude of \san1237 can be
produced by the combination of a giant branch star and a main-sequence
star close to the turnoff (Janes \&\ Smith 1984). No satisfactory
solution has been suggested yet to explain the magnitude and colour
of \san1072 (see e.g.\ Mathieu et al.\ 1990).

It is not clear to us why these three eccentric binaries emit X-rays.
We don't think that the X-ray emission is due to hot white-dwarf
companions, because the X-ray spectrum is not as soft
as would be expected for a white dwarf. Also, the progenitor of the
white dwarf would likely have circularized the orbit in its giant stage:
even an orbit of several thousand days can be circularized by
a white dwarf progenitor, as witnessed by \san1221 (Verbunt \&\ Phinney
1995).

\subsubsection{No indication of binarity}

A number of X-ray sources can be identified with stars for which there
is no indication that they are binaries. 
It cannot be excluded that these stars are chance coincidences.
\x43 $=$ \san1270 and \x44 $=$ \san775 have been observed by Latham
\&\ Mathieu; no radial velocity variation was found, so these stars
are not close binaries.
\x28 $=$ \san1112 has not been measured by Latham \&\ Mathieu.
\x35 is \san628, located to the left of the main-sequence; 
such a location can arise when a main-sequence star is accompanied
by a hot white dwarf.  The hardness of the X-ray spectrum
argues against a hot non-accreting white dwarf. Perhaps \san628 is a
cataclysmic variable. It should be noted that its membership probability is
only 69\%; if it is at larger distance than \m67 (to be at or above the
main sequence it has to be $\gtap 2.5$ times the distance of \m67),
its X-ray luminosity is accordingly higher.

\x4 is the blue straggler \san1082. This has been suggested to be
in a binary with a sdO companion on the basis of a large ultraviolet
excess with respect to the spectrum derived from Str\"omgren 
photometry (Landsman et al. 1998). \nocite{lbn+98}
A binary period of approximately 1 day has been suggested from optical
photometry
(Goranskij et al.\ 1992); \nocite{gkm+92} 
However, the large velocity variations implied by such a short period
have not been found by Mathieu et al.\ (1986), even though they note that
the dispersion in their velocities for this star is somewhat larger than 
normal. The larger dispersion may be
due to the early spectral type.\nocite{mlgg86}
\x19 is \san364, close to the location of the Mira variables in the Hertzsprung
Russell diagram. No chromospheric emission is present in \san364 
(Pasquini \&\ Belloni 1998).

\subsubsection{Non-members}

\x25 is identified with HD~75638. This star is a triple
star, with an inner orbit of 5.8$\,$d and zero eccentricity
(Nordstr\"om et al.\ 1997).  \nocite{nsla97} 
The X-ray luminosity suggests that the inner binary is a RS CVn 
system.

\x42 shows a flare in the last 2800$\,$s of our observations,
during which the countrate was 0.03 cts/s, which is about ten times
higher than the average countrate before.
Such a flare suggests that \san1042  may be a late main-sequence
star; its colour and magnitude are compatible with a K dwarf at the
distance of \m67.
However, according to Girard et al.\ (1989) \san1042 is not a member.

\nocite{ccc+90}
\begin{figure}
\centerline{\psfig{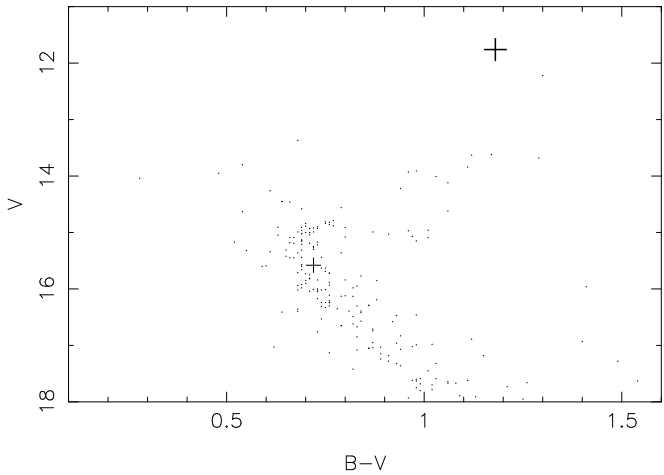} {\hfil}}
\caption[]{Hertzsprung Russell diagram of \ngc188 (data mainly from 
Caputo et al.\ 1990) showing as $+$ the two detected X-ray sources.
}
\label{n188}\end{figure}

\subsection{\ngc188 and V$\,$8}

In Fig.~\ref{n188} we show the Hertzsprung Russell diagram of \ngc188 with
the two detected member stars, \x29 and \x21.
\x29 $=$ D$\,$719 is one of the brightest giants in \ngc188, and
remarkable because of its relatively rapid rotation, at 
$v\sin i\simeq 24\,$km/s and Ca H and K emission.
The absence of radial-velocity variations suggest that it is a single,
rapidly rotating giant, an FK~Comae-type star
(Harris \&\ McClure 1985). \nocite{hm85}
The X-ray luminosity of the star, $L_{\rm 0.1-2.4 keV}=1.7\times 10^{31}\ergs$,
is in agreement with this suggestion.

High-precision radial-velocity measurements obtained with the WIYN telescope
show \x21 $=$ D$\,$1335 to be both a short-period velocity variable and a 
rapid rotator (Mathieu \&\ Dolan 1998, in preparation). 
Thus the X-ray emission at $L_{\rm 0.1-2.4 keV}=5.6\times 10^{30}\ergs$
is likely due to chromospheric activity. The star has been investigated for
photometric variability by Kaluzny \&\ Shara
(1987), who do not find it variable. \nocite{ks87}

\x26 $=$ D$\,$1361
is variable V$\,$8 of Kaluzny \&\ Shara (1987), who find evidence
for a 2.66$\,$day periodicity in the $B$ magnitude, and suggest
that the star is of the FK$\,$Comae type.
The proper motion of this star suggests that it is not a member
of \ngc188 (Dinescu et al. 1996).
If it is a member, its X-ray luminosity is $1.8\times 10^{32}\,$erg/s.

\section{Conclusions}

The increased sensitivity of the X-ray observation of \m67 and
the improved optical information have led to a true plethora
of source types, as illustrated by Table~3. 
Many of these are more or less expected source types, such as the
cataclysmic variable, the hot white dwarf, the contact binaries,
and the circular short-period binaries: for all of these the
X-ray luminosity is in the range found for similar sources
in the galactic disk. Interestingly however, several sources have
been detected for which we do not understand the mechanism
of X-ray emission. These include several binaries with eccentric
orbits, of which two are located in the Hertzsprung Russell
diagram below the subgiant branch.
Unexpected also are the detection of a binary of a giant and
a white dwarf in a circular orbit; and of a blue straggler
that appears to be a single star.

Further optical studies of these objects may help in elucidating the
mechanism(s) of X-ray emission. 
For example, chromospheric activity can be detected in the
form of H$\,\alpha$ or Ca~H and K emission; and is probably
caused by rapid rotation which may be detected through line broadening.
If mass transfer is taking place in any of these binaries, 
broad Balmer emission lines are expected.

The number of X-ray sources detected in \ngc188 is much lower,
due to the higher detection flux limit. One of the two detected
members is a single, rapidly rotating giant star, an FK Com type
object -- which adds yet another source type.
We predict that a more sensitive X-ray observation of \ngc188
will detect a few dozen X-ray sources, similar to those
we have found in \m67.
\nocite{san62}

\acknowledgements

This work builds on the Center for Astrophysics \m67 binary survey,
for which we express our appreciation to Dr.\ David Latham.
RDM was supported by a National Science Foundation grant 
AST-9731302. In identifying our X-ray sources
we have made use of the Open Cluster DataBase of Mermilliod (1996).

%\bibliographystyle{aabib}
%\bibliography{refs}

\end{document}